\documentclass[pre,aps,twocolumn,showpacs]{revtex4}
\usepackage{graphicx,epstopdf}
\usepackage{pdfpages}
\usepackage{dcolumn}
\usepackage{bm}
\usepackage{color}
\usepackage{amssymb,amsmath}
\usepackage [latin1]{inputenc}
\begin{document}

\title{Polymer escape through a three dimensional Double-Nanopore System}

\author{Swarnadeep Seth}
\author{Aniket Bhattacharya}

\altaffiliation[]
{Author to whom the correspondence should be addressed}
{}
\email{AniketBhattacharya@ucf.edu}

\affiliation{Department of Physics, University of Central Florida, Orlando, Florida 32816-2385, USA}
\date{\today}
\begin{abstract}
  We study escape dynamics of a double-stranded DNA (dsDNA) through an idealized double nanopore (DNP) geometry subject to two equal and opposite forces (tug-of-war) using Brownian dynamics (BD) simulation. In addition to the geometrical restrictions imposed on the cocaptured dsDNA segment in between the pores, the presence of tug-of-war forces at each pore results in a variation of the local chain stiffness for the segment of the chain in between the pores which increases the overall  stiffness of the chain.  We use BD simulation results to understand how the intrinsic chain stiffness and the TOW forces affect the escape dynamics by monitoring the local chain persistence length $\ell_p$, the residence time of the individual monomers $W(m)$ in the nanopores, and the chain length dependence of the escape time $\langle \tau \rangle$ and its distribution. Finally, we generalize the scaling theory for the unbiased single nanopore translocation for a fully flexible chain for the escape of a semi-flexible chain through a DNP in presence of TOW forces. We establish that the stiffness dependent part of the escape time is approximately independent of the translocation mechanism  so that $\langle \tau \rangle \sim \ell_p^{2/D+2}$, and therefore the generalized escape time for a semi-flexible chain can be written as 
$\langle \tau \rangle = AN^\alpha\ell_p^{2/D+2}$.  We use BD simulation results to compare the predictions of the scaling theory. Our numerical studies supplemented by scaling analysis provide fundamental insights to design new experiments where a dsDNA moves slowly through a series of graphene nanopores.
\end{abstract}
\maketitle
\section{Introduction}Recently  a double nanopore (DNP) platform has been suggested to be more effective alternative for analyzing DNA barcodes compared to the original design of single nanopore (SNP) and nanochannel based techniques~\cite{Reisner-PNAS-2010}. 
Unlike traditional methods, which require amplification, in a SNP)~\cite{Review}, or in a DNP platform~\cite{TwoPore} a particular DNA segment can be  analyzed while a dsDNA passes through the nanopore. 
Since its original demonstration in
$\alpha$-hemolysin protein pore~\cite{Kasianowicz,Meller00,Meller01,Meller02}, NP translocation has been
studied in other biological NPs, Silicon nanopores, and
multi-layered graphene NPs. In a DNP, compared to a single NP, a DNA segment between two tags can be analyzed multiple
times~\cite{Flossing} by keeping the segment captured in both the pores, resulting an increase in the accuracy of this method significantly. Moreover,
adjustable biases and feedback mechanism at each pore offer overall better control of the DNA. Different variations of this
concept, such as, two pores of different width~\cite{Dekker}, DNP separated by a nano-bridge~\cite{Cadinu1}, double-barrel NP~\cite{Cadinu2}, and
entropy driven TOW~\cite{Yeh} have been reported. \par
\begin{figure}[ht!]
\includegraphics[width=0.43\textwidth]{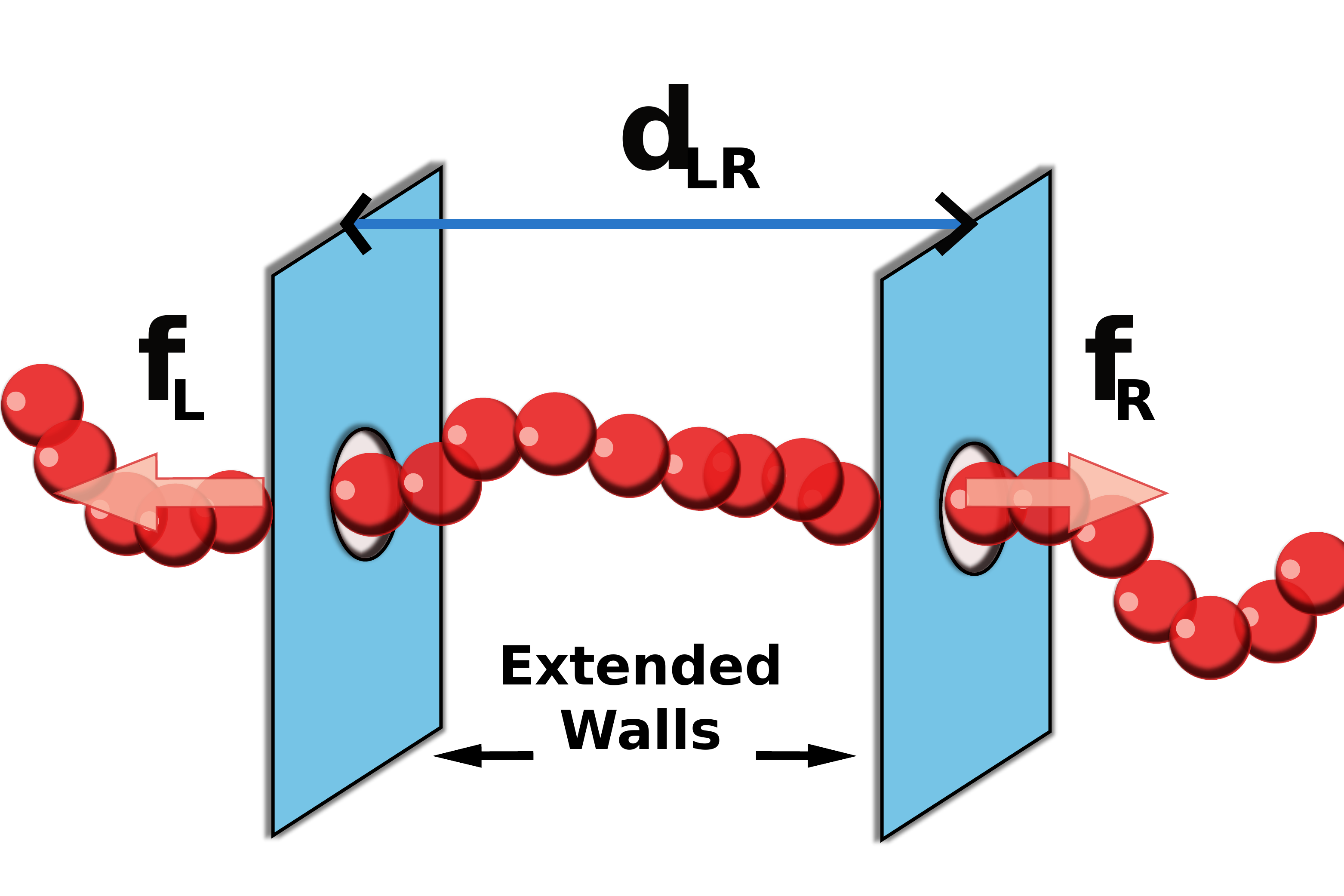}
\caption{\label{Model} Schematics of a chain of contour length $L$, where $\sigma$ is the diameter of the individual monomers cocaptured by the two nanopores separated by a distance $d_{LR}$. The two walls extend in $yz$ plain and external biases
  $\vec{f}_L=-|f_L|\hat{x}$ and $\vec{f}_R=|f_R|\hat{x}$  along negative and positive $x$ directions respectively are applied inside each pore of equal diameters $d_{pore}=2\sigma$.}
\end{figure}

While translocation through a 
SNP system has been studied quite extensively theoretically, experimentally,
and using a variety of numerical and simulation strategies~\cite{Review}, theoretical
studies and modeling translocation in double or multiple NP system is only just beginning~\cite{Bhattacharya_Seth_2020}.
In this letter,  we report  BD simulation
studies of a homopolymer escape through a DNP
system. The design of our ideal DNP system {\em in silico} (Fig.~\ref{Model}) has been motivated by recent experiment where the nanopores are drilled onto a single wafer as reported recently~\cite{Dekker,TwoPore,Flossing}, but our geometry resembles  a multilayered graphene nanopores, where first principles
transport calculations for DNA bases surveyed across a graphene nanopore system have 
illustrated the advantages of this geometry~\cite{Towfiq}. Thus, it is conceivable that future experiments will be carried out in this geometry of parallelly stacked graphene nanopores. 
Another purpose of choosing this geometry is that in the limit $d_{LR}/L << 1$, it is expected that some characteristics of the DNP translocation will show similarities with the corresponding quantities in the SNP translocation and can be analyzed 
using the known results
from scaling theory of SNP translocation~\cite{Ikonen_EPL2013,Ikonen_JCP2012,Ikonen_PRE2012}, nonequilibrium tension propagation (TP)
theory of polymer translocation~\cite{Sakaue_PRE_2007}, and prior results for SNP translocation for a stiff chain
~\cite{Adhikari_JCP_2013}. These studies will also provide information 
to design new experiments with different parameter sets, develop a theoretical framework
that can be tested by additional simulation studies.
\section{Model}
\label{Model0}
 Our BD scheme is implemented on a bead-spring model of a polymer with
 the monomers interacting via an excluded volume (EV), a Finite Extension Nonlinear Elastic (FENE) spring potential, and a
 bond-bending potential enabling variation of the chain persistence length $\ell_p$ (Fig.~\ref{Model}).  
The model,  originally introduced for a fully flexible chain by  Grest and Kremer~\cite{Grest}, has been studied quite 
extensively by many groups using both Monte Carlo (MC) and various molecular dynamics (MD) methods~\cite{Binder_Review}. 
Recently we have generalized the model for a semi-flexible chain and studied both equilibrium and dynamic properties~\cite{Adhikari_JCP_2013,Huang_EPL_2014a,Huang_JCP_2014}. 
Comparison of our BD results with those obtained for very 
large self-avoiding chains on a square lattice reveals robustness of the model for certain universal aspects, {e.g.}, 
scaling of end-to-end distance and transverse fluctuations~\cite{Huang_EPL_2014a, Huang_EPL_2014b,Huang_JCP_2014,Huang_JCP_2015}. 
Using our BD scheme for confined stiff polymers in nanochannels we have demonstrated and verified the existence of Odijk deflection length 
$\lambda \sim (\ell_pD^2)^{1/3}$~\cite{Huang_JCP_2015}. Last but not the least we have
used the same model earlier to address various problems in SNP
translocation with
success~\cite{Kaifu_PRL,Bhattacharya_EPJE,Bhattacharya2010,Bhattacharya_Proceedia}. The
successes of these prior studies explaining a variety of phenomena provide
assurance that the BD simulation studies will provide useful informations and insights toward a fundamental
understanding of polymer translocation through a model DNP system. 
\par
The EV interaction between any two monomers is given by a short range Lennard-Jones (LJ) potential
\begin{eqnarray}
U_{\mathrm{LJ}}(r)&=&4\epsilon \left[{\left(\frac{\sigma}{r}\right)}^{12}-{\left(\frac{\sigma}
{r}\right)}^6\right]+\epsilon, \;\mathrm{for~~} r\le 2^{1/6}\sigma; \nonumber\\
        &=& 0, \;\mathrm{for~~} r >  2^{1/6}\sigma.
\label{LJ}
\end{eqnarray}
Here, $\sigma$ is the effective diameter of a monomer, and 
$\epsilon$ is the strength of the LJ potential. The connectivity between 
neighboring monomers is modeled as a FENE spring with 
\begin{equation}
U_{\mathrm{FENE}}(r_{ij})=-\frac{1}{2}k_FR_0^2\ln\left(1-r_{ij}^2/R_0^2\right).
\label{FENE}
\end{equation}
Here, $r_{ij}=\left | \vec{r}_i - \vec{r}_j \right|$ is the distance
between the consecutive monomer beads $i$ and $j=i\pm1$ at $\vec{r}_i$
and $\vec{r}_j$, $k_F$ is the spring constant and $R_0$
is the maximum allowed separation between connected monomers. 
The chain stiffness $\kappa$ is introduced by adding an angle dependent three body interaction term between successive bonds 
as (Fig.~\ref{Model}) 
\begin{equation}
U_{\mathrm{bend}}(\theta_i) = \kappa\left(1-\cos \theta_i\right) 
\end{equation}
Here $\theta_i$ is the angle between the bond vectors 
$\vec{b}_{i-1} = \vec{r}_{i}-\vec{r}_{i-1}$ and 
$\vec{b}_{i} = \vec{r}_{i+1}-\vec{r}_{i}$, respectively, as shown in Fig.~\ref{Model}. The strength 
of the interaction is characterized by the bending rigidity $\kappa$
associated with the $i^{th}$ angle $\theta_i$.
For a homopolymer chain the bulk persistence length $\ell_p$ of the chain in
three dimensions (3D) is
given by~\cite{Landau}
\begin{equation}
  \ell_p/\sigma = \kappa/k_BT.
  \label{lp_bulk}
  \end{equation}
\par 
Each of the two purely repulsive walls consists of one mono-layer
(line) of immobile LJ particles of the same diameter $\sigma$ of the
polymer beads symmetrically placed at $\pm \frac{1}{2}d_{LR}$. 
The two nanopores are created by removing particles at the center
of each wall (Fig.~\ref{Model}). 
We use the Langevin dynamics with the following equations of motion for the i$^{th}$ monomer 
\begin{equation}
m \ddot{\vec{r}}_i = -\nabla (U_\mathrm{LJ} + U_\mathrm{FENE} + U_\mathrm{bend} \\
                                            + U_\mathrm{wall}) -\Gamma \vec{v}_i + \vec{\eta}_i . 
                                          \label{langevin}                                          
\end{equation}

Here $\vec{\eta} _ i (t)$ is a Gaussian white noise with zero mean at temperature $T$, and 
satisfies the fluctuation-dissipation relation in $d$ physical
dimensions (here $d=3$):
\begin{equation}
< \, \vec{\eta} _ i (t) 
\cdot \vec{\eta} _ j (t') \, > = 2dk_BT \Gamma \, \delta _{ij} \, \delta (t 
- t ').
\end{equation}
We express length and energy in units of $\sigma$ and $\epsilon$, respectively. 
The parameters for the FENE potential in Eq.~(\ref{FENE}), $k_F$ and 
$R_0$, are set to $k_F = 30 \epsilon/\sigma$ and $R_0 = 1.5\sigma$, respectively. 
The friction coefficient and the temperature are set to 
$\Gamma = 0.7\sqrt{m\epsilon/\sigma^2}$, $k_BT/\epsilon = 1.2$,
respectively. The force is measured in units of $k_BT/\sigma$.

The numerical integration of Equation~(\ref{langevin}) is implemented using the algorithm introduced by Gunsteren and Berendsen~\cite{Langevin}.   Our previous experiences with BD simulation suggests that for a time step $\Delta t = 0.01$ these parameters values produce stable trajectories over  a very long period of time and do not lead to unphysical crossing of a bond by a monomer~\cite{Huang_JCP_2014,Huang_JCP_2015}.  The average bond length stabilizes at $b_l = 0.971 \pm 0.001$ with negligible fluctuation regardless of the chain size and rigidity~\cite{Huang_JCP_2014}. We have used a Verlet neighbor list~\cite{Allen} in stead of a link-cell list to expedite the computation.
\section{Results}
The starting conformation of our BD simulation is a DNA polymer already captured and threaded through both the pores as in Fig.~\ref{Model}. We symmetrically place the polymer in a DNP device and equilibrate the polymer chain keeping two polymer beads inside each pore clamped. We equilibrate the polymer over $10$ times the Rouse relaxation time $\tau_{Rouse} \sim N^{1+2\nu}$, where $\nu =0.588$ is the Flory exponent in 3D (for N=128 this corresponds to 10$^8$ time steps)~\cite{Rubinstein}, the polymer chain is allowed to translocate under the influence of two external forces $\vec{f}_L$ and $\vec{f}_R$. In this paper we only consider the TOW $\vec{f}_L+ \vec{f}_R=0$, so that the polymer chain diffuses across the entropic barrier imposed by the pores.  To calculate relevant physical quantities, we take average over $2000$ successful translocation events for several chain lengths $L=128\sigma - 256\sigma$ and for several values of stiffness parameter $\kappa = 0 - 128$. 
\subsection{Chain persistence length during translocation:}
\begin{figure}[ht!]
  \includegraphics[width=0.45\textwidth]{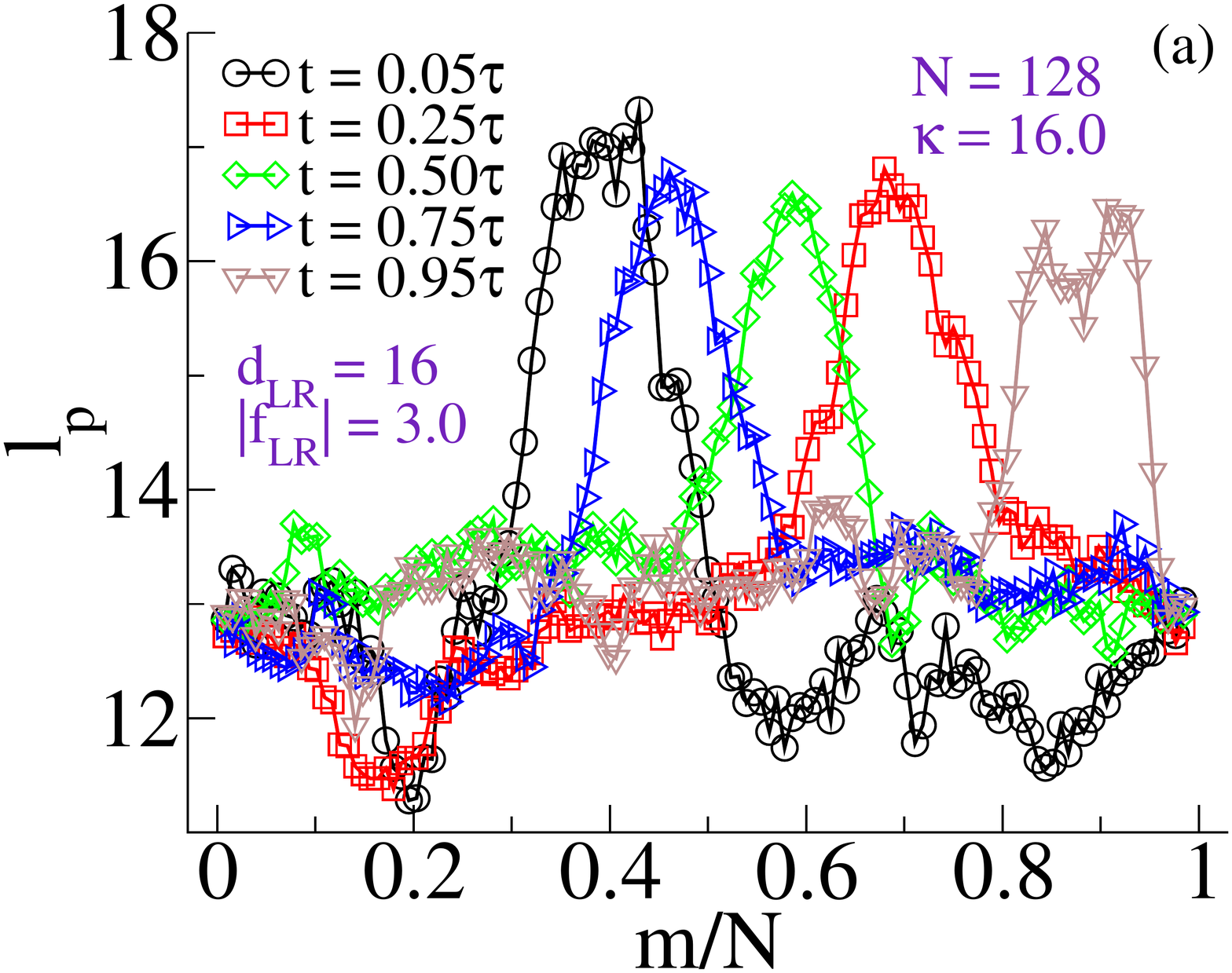}\par
  \vskip 0.125truecm
\includegraphics[width=0.47\textwidth]{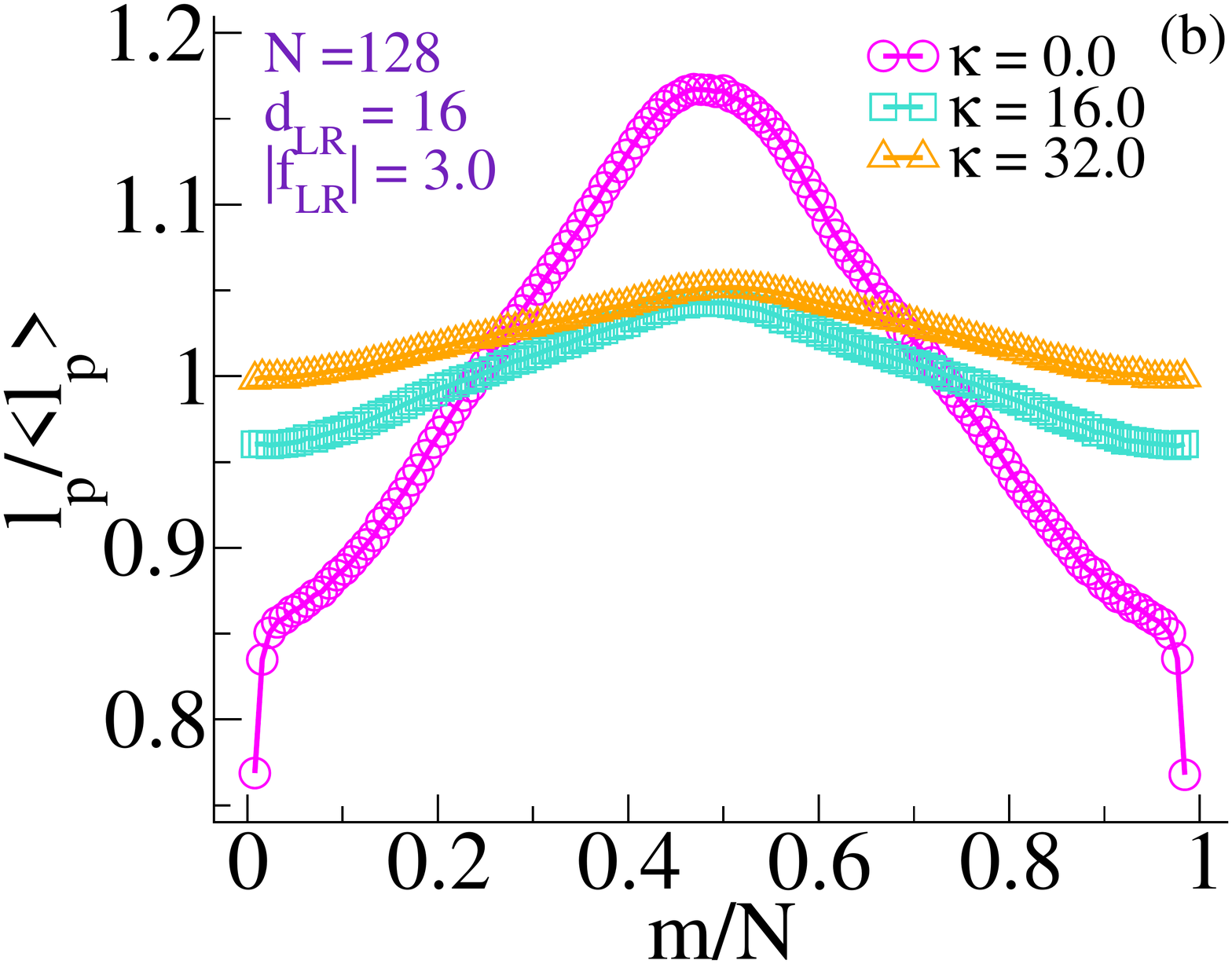}
\caption{\label{lp}\small (a) The instantaneous  local chain persistence length $\ell_p(m)$  at five different instances
($t=0.05\tau$ (black circles), $t=0.25\tau$ (red squares), $t=0.50\tau$ (green diamonds), $t=0.75\tau$ (blue triangles), and $t=0.95\tau$ (brown down triangles) showing that different parts of the chain become stiffer at different time of the translocation process. (b) Normalized time averaged persistence length $l_p/\langle\ell_p \rangle$ as a function of reduced monomer index $m/N$ for $\kappa=0$ (magenta circles), $\kappa=16$ (teal squares) and $\kappa=32$ (orange triangles) respectively. The relative increase in $\ell_p$ is most prominent for $\kappa = 0$.}
\end{figure}

The conformations of the chain segment in between the pores are severely restricted. In addition, opposite forces are present at each pore. This provides the chain segment a time dependent stiffness as shown in Fig.~\ref{lp}(a).  Here we show the instantaneous local chain stiffness $\ell_p(m)$ as a function of the monomer index $m$ for one successful translocation event at times 0.05$\tau$, 0.25$\tau$, 0.5$\tau$, 0.75$\tau$ and 0.95$\tau$, $\tau$ being the total translocation time. Fig.~\ref{lp}(a)
shows that the chain segment acquires an increased stiffness while crossing the region in between the two pores. As time progresses, the position of the maxima does not necessarily
occur at an increasing value of the reduced monomer index $m/N$ due to back and forth motion of the chain.  
The restricted motion of the monomers in the vicinity of pores and the presence of TOW forces make the chain locally stiffer. We also calculated the time averaged stiffness shown in Fig.~\ref{lp}(b). Since 
$\vec{f}_L+ \vec{f}_R=0$, the middle monomers spend considerable amount of time, as also reflected in the wait time distribution $W(m)$ (please see next section)  in the region in between the pores resulting the average stiffness becoming maximum around $m=N/2$. 
While calculating the local persistence length we have used the same expression as that of a worm-like chain (WLC)~\cite{Rubinstein}
\begin{equation}
  \ell_p(m) = -\frac{1}{\ln[\cos(\theta_m)]},
  \label{lp_eqn}
  \end{equation}
  where $\theta_m$ is the angle subtended by the adjacent bond vectors connecting the monomer $m$.
  Previously we have shown that the inclusion of the excluded volume interaction does not alter Eqn.~\ref{lp_eqn}, as this is a
  local property of the chain~\cite{Huang_JCP_2014}.
  It is also evident from Fig.~\ref{lp}(b) that the relative increase in persistence length $\ell_p/\langle \ell_p \rangle$ is most significant for $\kappa=0$ 
and becomes less prominent for a stiffer chain with higher $\kappa$ value which resembles the stiffness of a double-stranded DNA . The chain segment in between the pores experiences equal and opposite forces which further increases the stiffness by restricting entropic penalty. This reduction is entropy is less significant for a stiffer chain which explains the effect.  
\subsection{The Wait time Distribution:~} 
\begin{figure}[ht!]
\includegraphics[width=0.45\textwidth]{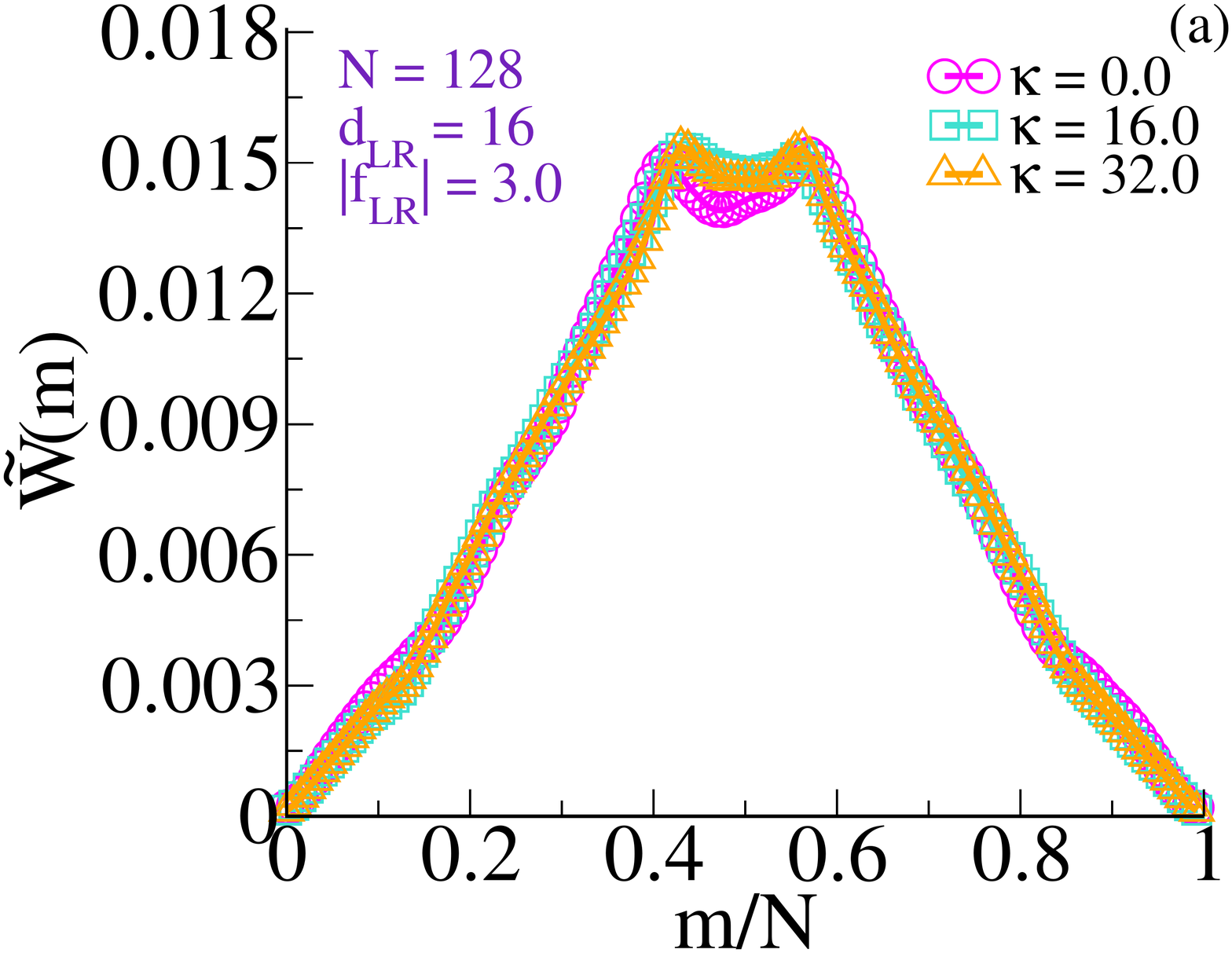}\par
\vskip 0.125truecm
\includegraphics[width=0.43\textwidth]{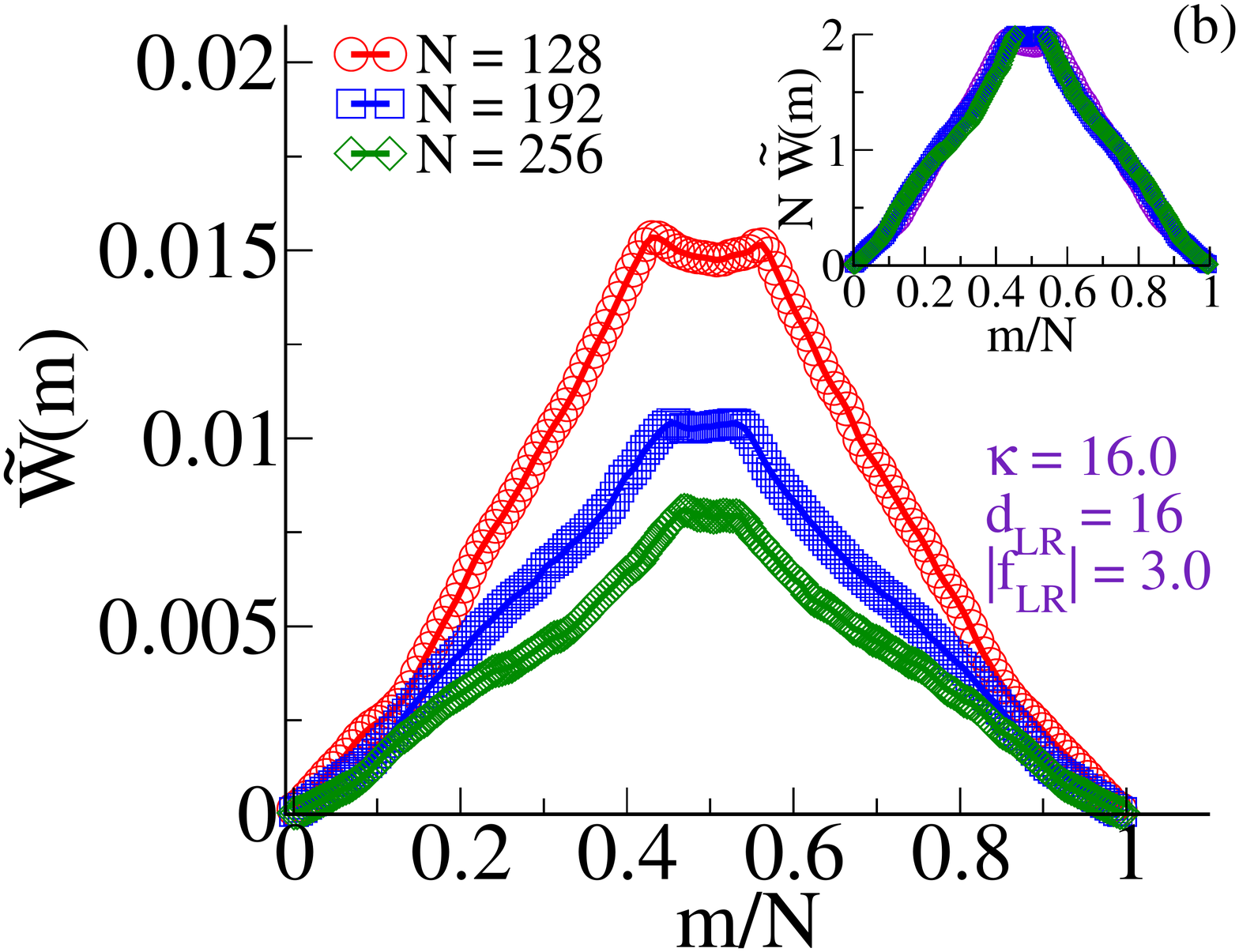}
\caption{\label{wait_time}\small (a) Normalized wait time distribution $\tilde{W}(m)$ as a function of reduced monomer index $m/N$ for $N=128$ for
  $\kappa =0$ (magenta circles), 16 (teal squares), and  32 (orange diamonds) respectively.  (b) $\tilde{W}(m)$ shows the chain length dependence ($N=128$, 192, and 256 of $W(m)$ for $\kappa =16.0$. The inset confirms that qualitative behavior remains uniform for $N\tilde{W}(m)$ with reduced monomer index for different chain lengths.}
\end{figure}

A useful quantity that provides the detail of the translocation process is the residence time or the wait time distribution $W(m)$ of the translocating chain. The normalized $\tilde{W}(m)$ is defined as 
\begin{equation}
\langle \tilde{W}(m) \rangle = \frac{1}{\sum_{m=1}^N W(m) }{\langle W(m) \rangle}.
\end{equation}
By definition $\sum_{m=1}^N W(m)=\tau$, hence  $\sum_{m=1}^N \tilde{W}(m)=1$. Thus $\tilde{W}(m)$ provides the relative time spent by the  individual monomer during the translocation process as shown in Fig.~\ref{wait_time}. The quantity has been calculated for the unbiased SNP translocation and shares qualitatively similar features~\cite{Luo_JCP2006} excepting minor modification in its shape near $ N/2 - \frac{1}{2}d_{LR}/\sigma   < m < N/2 + \frac{1}{2}d_{LR}/\sigma $ due to the presence of two pores.
For the unbiased translocation in a SNP, the $W(m)$ is symmetric and peaks at $m=N/2$,  simply due to the fact the entropic force is balanced at either side of the pore, as has been observed previously~\cite{Luo_JCP2006}. For the escape problem in a DNP in a TOW situation, $W(m)$ is still symmetric around $m=N/2$ but now the two peak positions shift to  $m \simeq N/2 \pm \frac{1}{2}d_{LR}/\sigma$ for the exact same reason as for this shape the entropic forces are balanced at the left side of the left pore and at the right side of the right pore. Similar to what is observed for the SNP, $\tilde{W}(m)$ rises roughly linearly  for $m< N/2 - \frac{1}{2}d_{LR}/\sigma$, 
until peaks at  $m \simeq N/2 - \frac{1}{2}d_{LR}/\sigma$. It then
decreases to a minimum at $m =N/2$, rises and peaks again at $m =  N/2 + \frac{1}{2}d_{LR},/\sigma$, then goes down
almost linearly as shown in Fig.~\ref{wait_time}(a). The two noticeable kinks at $0.5 \pm 0.3$, where a change of slope occurs, are when the monomers have exited either of the pores and subject to a net bias force. It is also worth noticing that
$W(m)$ has a local minimum at the midpoint of the chain $m=N/2$.  The monomer with index $m=N/2$ lowers the free energy by staying equidistant from the two pores which decreases its residence time at each pore. This explains the shape of Fig.~\ref{wait_time}(a).  We further observe that $\tilde{W}(m)$ is almost insensitive to the chain stiffness.
An increase in the chain stiffness causes the translocation time $\langle \tau \rangle$ to increase ~\cite{Adhikari_JCP_2013} (shown at the inset) and collapse of $\langle \tilde{W}(m) \rangle $ for different stiffness onto the same master curve  
implies that $\langle \tilde{W}(m) \rangle $ for each $m$ increases proportionally with the translocation time. In addition, Fig.~\ref{wait_time}(b) confirms $\langle \tilde{W}(m) \rangle$ scales uniformly with chain lengths and the inset shows an excellent data collapse for $N\langle \tilde{W}(m) \rangle$ against reduced monomer index. \par
\begin{figure}[ht!]
\includegraphics[width=0.45\textwidth]{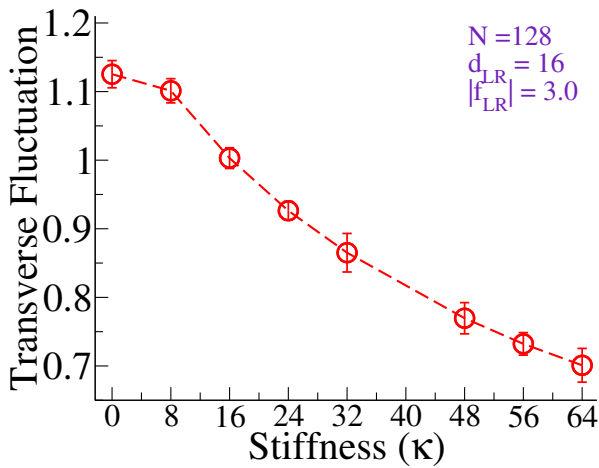}
\caption{\label{fluct}\small Transverse fluctuation shows a steady decrease with increasing chain stiffness $k$. }
\end{figure}

Interestingly, we observe that $\tilde{W}(m)$ is also insensitive to the rms transverse fluctuation $\sqrt{\langle l_{\perp}^2\rangle}$  of the segment in between the pores that decreases monotonically with increasing chain stiffness (Fig.~\ref{fluct}) .
In simulation we measure the rms transverse fluctuation as follows:
\begin{equation}
   \sqrt{ \langle l_{\perp}^2 \rangle}
   = \sqrt{ \frac{1}{m_{pore}} \sum_{i_{pore}=1}^{m_{pore}} \left( y_i^2 + z_i^2 \right)} ,
\end{equation}
where $y_i$ and $z_i$ are the vertical distances of the i$^{th}$ monomer with respect to the direction $\hat{x}$ from the left pore the right pore, and $m_{pore}$ are the number of monomers in between the two pores.
This decrease in $\sqrt{\langle l_{\perp}^2\rangle}$ is a generic feature for chain segment under tension. We have explained it elsewhere by mapping the translocation problem to that of a {\em flexible-stiff-flexible} triblock copolymer~\cite{Bhattacharya_Seth_2020}. Combining results from Figs. 2, 3, and 4 we conclude that despite variations in chain stiffness and concomitant fluctuations in between the pores, the normalized wait time distributions $\tilde{W}(m)$ collapse onto the same master curve.  This result can be helpful to explain characteristics of translocating polymers of different stiffness and contour length.
\subsection{The TOW and the Escape:~}
A central question in polymer translocation is how long does it take for the chain to escape from one side to the other~? This has been described in terms of a translocation exponent $\alpha$ which determines the power law dependence of the mean translocation time $\langle \tau \rangle$ on the chain length $N$,
\begin{equation}
  \langle \tau \rangle = AN^\alpha.
  \label{alpha}
\end{equation}
We expect that for $d_{LR} << L$, the escape problem in a DNP will also be described by a similar power law dependence of the $\langle \tau \rangle$ on $N$. In this limit the contribution of the segment  in between the two nanopores becomes insignificant, thus, one expects that the exponent $\alpha$ will be the same as that of a SNP.
We have obtained the escape time 
$\langle \tau \rangle$  from the BD simulation for the symmetrically placed and cocaptured polymer in between two nanopores for $\vec{f}_L + \vec{f}_R = 0$.  Due to the symmetric arrangement,  the case escape occurs equally through the left pore and the right pore. By monitoring 2000 independent runs we checked that  translocations occur with equal probability in either direction. \par
The escape problem through an idealized nanopore in a thin membrane has been studied theoretically by various authors~\cite{Muthukumar,Sung_Park,CKK}.
For a fully flexible chain it was proposed ~\cite{Muthukumar,CKK} that $\alpha = 1+2\nu$ so that  
the average translocation time for the unbiased translocation scales with the chain length as
$\langle \tau_0 \rangle = AN^{1+2\nu}$. 
This follows assuming quasi equilibrium condition so that the gyration radius
$\langle R_g \rangle \sim N^\nu$, is the same as that of the bulk, and that in absence of the hydrodynamic effects, the diffusion constant of the center of mass of the chain $D \sim 1/N$. Thus, translocation time $\langle \tau \rangle$ to travel  a distance of the order of $\langle R_g\rangle$ can be estimated by substituting
\begin{equation}
  \langle R_g^2\rangle \sim N^{2\nu},
  \label{Flory}
\end{equation}
 in the diffusion equation $\langle R_g^2 \rangle \sim D\langle \tau \rangle$, which results
\begin{equation}
  \label{ckk}
 \langle \tau_0 \rangle \sim \langle R_g^2 \rangle/D = A N^{1+2\nu}.
\end{equation}
Thus, the translocation time in this picture scales as the Rouse relaxation time~\cite{Rubinstein}.\par
\begin{figure}[ht!]
\includegraphics[width=0.45\textwidth]{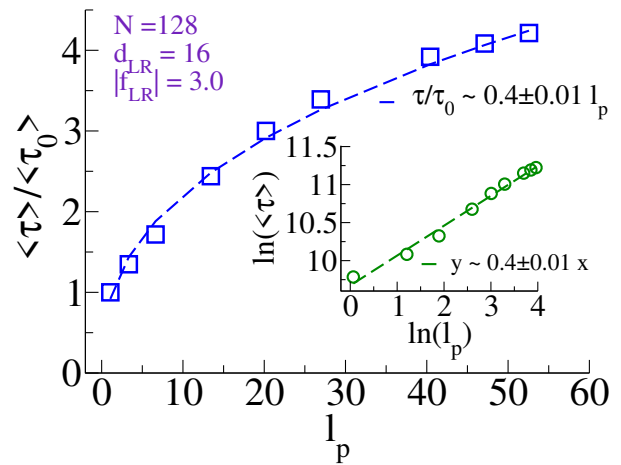}
\caption{\label{tau2}\small $\langle \tau_{\ell_p} \rangle/\langle \tau_0 \rangle $ as a function of $\ell_p$ (blue squares). The dotted line is a
  power law fit  ($\langle \tau_{\ell_p} \rangle/\langle \tau_0 \rangle \sim \ell_p^{0.4} $)   through the points validating Eqn.~\ref{tau_lp1}. The inset 
  shows the same (green circles) on a log-log scale. The straight line is a linear fit with slope $0.40\pm 0.01 $. }
\end{figure}

Eqn.~\ref{ckk} for a fully flexible chain can be generalized for a semiflexible chain using the generalized Flory theory due to Nakanishi~\cite{Nakanishi} and Schaeffer,
Pincus and Joanny~\cite{Pincus_MM_1980} which incorporates the persistence length $\ell_p$ into the Flory Eqn.~\ref{Flory} as follows.
\begin{equation}
  \sqrt{\langle R_g^2\rangle} \sim \ell_p^{1/(D+2)}N^\nu
  \label{Nakanishi}
\end{equation}
in $D$ physical dimensions. Previously we
have shown that in two dimensions (2D) 
Eqn.~\ref{Nakanishi} holds for $L/\ell_p > 1$~\cite{Huang_JCP_2014}.
Hence for the unbiased translocation the generalization for the translocation time  $\langle \tau_{\ell_p} \rangle$ for a semi-flexible chain of persistence length $l_p$ is
\begin{equation}
 \langle \tau_{\ell_p} \rangle = A  \ell_p^{\frac{2}{D+2}}N^{1+2\nu}=
\langle \tau_0 \rangle \ell_p^{\frac{2}{D+2}}
\label{tau_lp1}
\end{equation}
In making generalization of Eqn.~\ref{ckk} to Eqn~\ref{tau_lp1} we assumed that the amplitude factor $A$ remains the same. In other words we have decoupled that stiffness factor from the intrinsic translocation time of a fully flexible chain, an assumption is not fully justified {\em a priori}. But we observe this works reasonably  well for the escape problem that we have studied here.
Slater~\cite{Slater_JCP2012} and Panja~\cite{Panja} has suggested an alternative expression for $\langle \tau_0 \rangle$ based on the memory effect to Eqn.~\ref{ckk} where the exponent $\alpha = 2+\nu$, so that $\langle \tau_0 \rangle \sim N^{2+\nu}$.
If we assume that the stiffness factor  $\ell_p^{2/D+2}$ decouples from the solvent factor then
instead of Eqn.~\ref{tau_lp1} one gets, 
$ \langle \bar{\tau}_{\ell_p} \rangle = \bar{A}  \ell_p^{\frac{2}{D+2}}N^{2+\nu}=\langle \bar{\tau}_0 \rangle \ell_p^{2/D+2}$
Here $\langle \bar{\tau}_0\rangle =\bar{A}N^{2+\nu}$.
We assume that the stiffness factor  $\ell_p^{2/D+2}$ enters in to this equation exactly the same way as in Eqns.~\ref{tau_lp1} irrespective of the mechanism of $\langle \bar{\tau}_0\rangle$.
Fig.~\ref{tau2} verifies this decoupling of the stiffness factor from the intrinsic translocation time $\langle \tau_0 \rangle$ for a fully flexible chain.
In 3D the factor $\ell_p^{2/D+2} = \ell_p^{0.4}$. A plot of $\langle \tau_{\ell_p}\rangle/ \langle \tau_0\rangle$ as a function of the chain persistence length $\ell_p$
  validates the the prefactor $\ell_p^{2/D+2} = \ell_p^{0.4}$ in Eqns.~\ref{tau_lp1}. In the next section we explore the chain length dependence of $\langle \tau_0\rangle \sim N^\alpha$ in the DNP system. \par 
The polymer escape problem has been studied by several authors in the past making its connection to the translocation
problem~\cite{Muthukumar,Sung_Park,CKK,Luo_JCP2006,Slater_PRE2009,Slater_JCP2012}. One would like to distinguish between translocation and escape in this context.
Typically, in a translocation problem, the entire chain crosses from the {\em cis} side of the pore to the {\em trans} side. For the case of driven translocation
simulation studies are also carried out by placing the 1st monomer either at the center of the pore or slightly shifted at the {\em trans}
side~\cite{Bhattacharya_Proceedia, Bhattacharya_EPJE,Bhattacharya2010}. 
For the unbiased case this will be prohibitively large, as the polymer has to cross a huge barrier. In order to circumvent this problem CKK put an artificial constraint that once a monomer is on the {\em cis} side, it can not go back to the {\em trans side}~\cite{CKK}. With this constraint their numerical calculation converged on the translocation time $\langle \tau \rangle = AN^{1+2\nu}$. CKK further argued that the prefactor is larger compared to the unconstrained case in order to account for the slower diffusion due to constraint imposed by the nanopore and the wall, and concluded that for the unbiased case the translocation exponent is the same as the as that of the relaxation process.\par
\begin{figure}[ht!]
\includegraphics[width=0.46\textwidth]{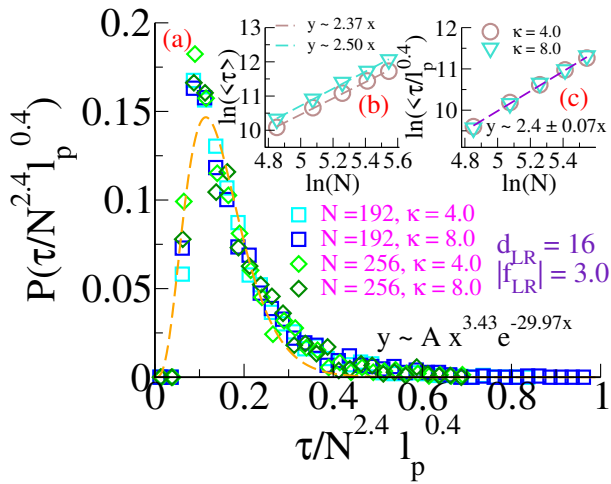}\par 
\caption{\label{MFPT_hist1}\small (a) Histograms of normalized escape time $\tau/N^{2.4}\ell_p^{0.4}$ for chain length $N = 192$
  (cyan and blue squares for $\kappa = 4.0$ and 8.0) and  for $N=256$  (light and dark green diamonds for $\kappa = 4.0$ and 8.0) respectively show data collapse. (b)  $\langle \tau \rangle$ as a function of $N$ (log scale) shows slopes 2.37 and 2.50 for  $\kappa = 4.0$ and 8.0 respectively. (c) The renormalized 
  $\langle \tau \rangle/\ell_p^{0.4}$ as a function of $N$ (log scale) which shows collapse of both the curves with slope 
  $\alpha = 2.40 \pm 0.07$ consistent with (a).}
\end{figure}
\begin{figure}[ht!]
  \includegraphics[width=0.46\textwidth]{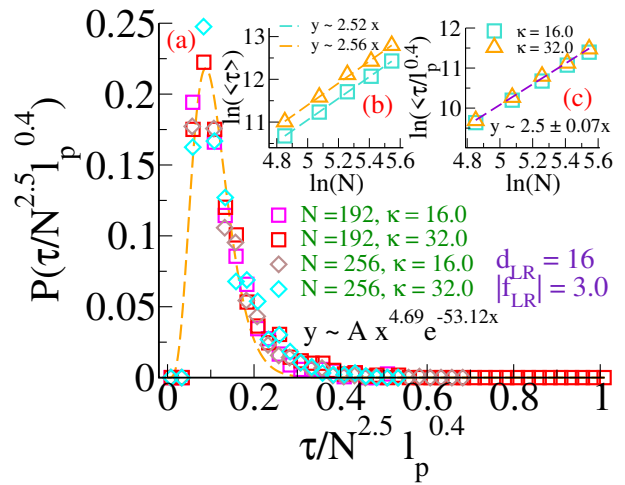}\par
\caption{\label{MFPT_hist2}\small (a) Histograms of normalized escape time $\tau/N^{2.5}\ell_p^{0.4}$ for chain length $N = 192$
  (magenta and red squares for $\kappa = 16.0$ and 32.0) and  for $N=256$  (brown and cyan diamonds for $\kappa = 16.0$ and 32.0) respectively show data collapse. (b)  $\langle \tau \rangle$ as a function of $N$ (log scale) shows slopes 2.52 and 2.56 for  $\kappa = 16.0$ and 32.0 respectively. (c) The renormalized 
  $\langle \tau \rangle/\ell_p^{0.4}$ as a function of $N$ (log scale) which shows collapse of both the curves with slope 
  $\alpha = 2.50 \pm 0.07$ consistent with (a).}
\end{figure}
 Dubbeldam {\em et al.}~\cite{Dubbeldam_PRE2007} mapped that escape problem in one-dimensional anomalous diffusion problem in terms of reaction coordinate and predicted the anomalous exponent $\alpha^{'} = 2/(2\nu +2 + \gamma_1) = 0.801 (3d)$ by introducing surface exponent term $\gamma_1 = 0.68 (3d)$.  Dubbeldam {\em et al.} set up MC simulation with a directional constraint on polymer movement and showed $\langle \tau_0 \rangle \propto N^{2/\alpha^{'}} = N^{2.496}$ which agrees their theoretical framework.
de Haan and Slater~\cite{Slater_JCP2012} incorporated memory effects during unbiased-translocation and asymptotically estimated $\langle \tau_0\rangle \sim N^{2.516}$ for a very long polymer.
\par 
Luo {\em et al.} revisited that same problem in two dimensions and studied polymer escape through a SNP using bond-fluctuation model~\cite{Luo_JCP2006}.
Their initial condition is the same as ours, namely, to release the polymer from the peak of the entropic barrier and let it diffuse down the entropic valley. They observed that  for the escape problem in 2D ($\nu = 0.75$)  $\langle \tau \rangle \sim N^{2.5}$ confirming that the escape problem and the translocation problem has the same (relaxation) exponent $1+2\nu$. \par

We studied the same problem albeit in the context of a DNP and having the DNA in a TOW with two equal and opposite forces so that net force is zero. Specifically, we studied the variation of the exponent $\alpha$, when one increases the chain persistence length from $\ell_p < d_{LR}$ to  $\ell_p > d_{LR}$.  The logarithmic
plots in Figs.~\ref{MFPT_hist1}(b) ($\ell_p < d_{LR}$) and \ref{MFPT_hist2}(b) $\ell_p > d_{LR}$ show a systematic increase in slope ($\alpha$) from 2.37 to 2.52 as expected due to the stiffness factor $\ell_p^{0.4}$ in Eqn.~\ref{tau_lp1} as $\ell_p$ is increased from 4 - 32. However, Figs.~\ref{MFPT_hist1}(c) and ~\ref{MFPT_hist2}(c) show plots of $\tau_{\ell_p}/{\ell_p}^{0.4}=\langle \tau_0\rangle \sim N^\alpha $ (see Eq.~\ref{tau_lp1}) where the data for different $\ell_p$ collapse on to the same straight line. We obtain
 $\alpha = 2.4 \pm 0.05$ for $\kappa = 4$ and 8 ($\ell_p < d_{lR}$) and
$\alpha = 2.5 \pm 0.04$ for $\kappa = 16$ and  32 ($\ell_p > d_{lR}$) respectively. This is further ensured from the  data collapse in
Figs.~\ref{MFPT_hist1}(a) and \ref{MFPT_hist2}(a). We further observe that as in the case of a DNP translocation problem, the shape of the histogram can be fitted with  $P(x) \sim x^{\alpha}\exp(-\beta x)$ with the maximum located at $\alpha/\beta$. We verify that
$\int_{0}^{\infty}x.P(x) dx$ returns back the mean translocation time $\langle \tau \rangle$. Thus, for the DNP translocation we observe that the translocation exponent $\alpha > 1+2\nu$ and $\alpha$ increases with increasing stiffness. Since we have decoupled the intrinsic chain stiffness, this slow down of the translocation process is likely due to additional constraint imposed by the DNP.
\section{Summary and Concluding Remarks}
We have studied the polymer escape problem in a DNP system in a TOW situation where the distance between the pores is much smaller than the chain contour length for several chain persistence lengths. The problem bears similarities with much studied problem of escape of a fully flexible chain through a SNP. However, because if the presence of the two equal and opposite forces, our simulation studies reveal additional intriguing features. First, during the escape process the chain segments in the region between the pores acquires increased stiffness which becomes more prominent for a fully flexible chain, but also noticeable for stiffer chains. In contrast, the scaled wait time distributions for different chain stiffness collapse on to the same master curve, indicating that an increased chain stiffness introduces a global shift in the wait time distribution of the individual monomers with respect to the total translocation time.\par

We proposed generalization of the chain length dependence of the escape problem for semi-flexible chains (Eqn.~\ref{tau_lp1}). Our simulation data establishes an important aspect of the escape problem that the stiffness factor arising from the generalization of the Flory theory (Eqn.~\ref{Nakanishi}) decouples from the escape time of a fully flexible chain and thus, to a first approximation the theories developed for a fully flexible chain can be applied here leading to Eqn.~~\ref{tau_lp1}. 
Our simulation studies validates  that for a given contour length the escape time increases as a power law $\ell_p^{2/D+2}$. For the unbiased translocation the chain conformations are in quasi-equilibrium and hence Eqn~\ref{Nakanishi} is valid. However, we observe from the plot of
$  \langle \tau \rangle /\ell_p^{2/D+2}=  \langle \tau_0 \rangle = AN^\alpha$ (Insets of Fig.~\ref{MFPT_hist1}(b) and Fig.~\ref{MFPT_hist2}(b)) the translocation exponent $\alpha$ depends on the chain stiffness and increases from $2.4 \pm 0.01$ to $2.5 \pm 0.01$. This arises due to the additional constraint imposed by the two pores.
and the system to develop chracateristics of a reptation~\cite{Rubinstein}, which is intrinsically a slower process than diffusion. 
We believe these results will be useful for future DNP experiments driven under low bias in dual graphene nanopores.
 \section{Acknowledgment}
The authors acknowledge computing resources under the auspices of
UCF's high performance computing cluster STOKES where all the
computations were done and thank Professor K. Binder and Professor W. Reisner for discussions.
\section{DATA AVAILABILITY}
The data that support the findings of this study are available from the corresponding author upon reasonable request.

\end{document}